\documentclass{mem}
\usepackage{natbib}\usepackage{txfonts}\usepackage{balance}
\usepackage{graphicx}
\usepackage[a4paper]{hyperref}
\idline{75}{282}
\begin{document}
\def\teff{$T\rm_{eff }$}
\def\kms{$\mathrm {km s}^{-1}$}
\newcommand{\et}{\textit{et al. }}
\newcommand{\hhh}{\mbox{H$_2$ }}
\newcommand{\HI}{\mbox{H\textsc{i}}\,\,}
\newcommand{\ma}{\mbox{m\AA}}

\title{Cosmological observations to shed light on possible  variations}

   \subtitle{expectations, limitations and status quo}

\author{
M. \,Wendt\inst{1} 
D. \,Reimers\inst{1}
\and P. \, Molaro\inst{2}
          }

  \offprints{M. Wendt}

\institute{
Hamburger Sternwarte, Universit\"at Hamburg, Gojenbergsweg 112, 21029 Hamburg,
Germany
\email{mwendt@hs.uni-hamburg.de}
\and
Osservatorio Astronomico di Trieste, Via G.\,B.\,Tiepolo 11, 34131 Trieste,
Italy
}

\authorrunning{M. Wendt}


\abstract{
Cosmology contributes a good deal to the investigation of variation of fundamental physical constants.
High resolution data is available and allows for detailed analysis over cosmological
distances and a multitude of methods were developed.
The raised demand for precision requires a deep understanding of the limiting
errors involved. The achievable accuracy is under debate and current
observing proposals max out the capabilities of todays technology.
The question for self-consistency in data analysis and effective techniques to
 handle unknown systematic errors is of increasing importance. 
An analysis based on independent data sets is put forward and 
alternative approaches for some of the steps involved are introduced.
}
\maketitle{}

\section{Introduction}
This work is motivated by numerous findings of different groups that partially are in
disagreement witch each other. A large part of these discrepancies reflect the different
methods of handling systematic errors. Evidently systematics are not yet under control
or fully understood. We try to emphasize the importance to take these errors, namely i.e.
calibration issues, into account and put forward some measures adapted to the problem.
\section{Data}
\subsection{Observations}
Q0347-383 is one of the few quasars with an absorption spectrum that shows
unblended strong features of molecular hydrogen. With a redshift of
$z_{\mathrm{abs}}=3.025$ the UV transitions are shifted into a spectral range
that can be observed with earthbound telescopes, rendering it an applicable
target for  
 $\Delta\mu/\mu$ analysis. Hence, Q0347 was subject of several works including
the paper by \cite{Reinhold06} that indicated a variation in $\mu$ and follow
up papers by \cite{King08}, \cite{Wendt08} and \cite{Thompson09} which
re-evaluated the data and report a result in agreement with no variation. All
recent works on Q0347 are based on the same UVES VLT
observations\footnote{Program ID 68.A-0106.} in January 2002 \citep{Ivanchik05}.
 For the first time additional observational data of Q0347 is taken into account
in this work. The data were acquired in 2002 at the same
telescope and recently reduced by Paolo
Molaro\footnote{Program ID 68.B-0115(A).}.
\subsection{Reduction}
The standard UVES pipeline may not reach the desired accuracy for precise
determination of fundamental physical constants. A major source of uncertainty
are the calibration spectra and the limited precision of the available list of
Thorium Argon lines as pointed out by \cite{Murphy08} and \cite{Thompson09}.
Proper calibration and data reduction will be the key to detailed analysis of
potential variation of fundamental constants. The influence of calibration
issues on the data quality is hard to measure and the magnitude of the resulting
systematic error is under discussion.
This work tries to bypass the estimation of systematics and utilizes the
uncertainties as part of the analysis procedure.
\begin{figure}[]
\resizebox{\hsize}{!}{\includegraphics[clip=true]{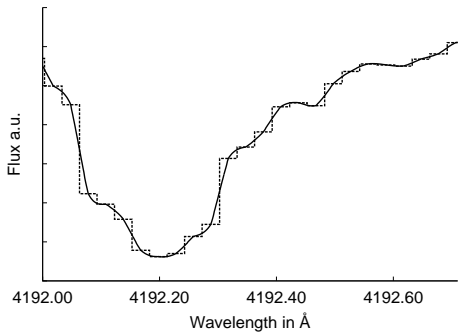}}
\caption{\footnotesize
The original flux is interpolated by a polynomial using {\it Neville's algorithm}
to conserve the local flux.}
\label{sampled}
\end{figure}
\begin{figure}[]
\resizebox{\hsize}{!}{\includegraphics[clip=true]{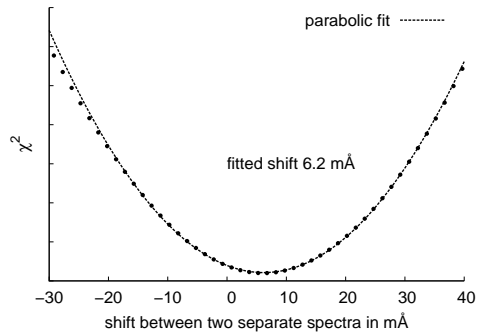}}
\caption{\footnotesize
Exemplary plot of the sub-pixel cross-correlation. The resulting shift is
ascertained
via parabolic fit. In this case the two spectra are in best agreement with a
relative shift of 6.2 \ma.}
\label{eta}
\end{figure}
\subsection{Preprocessing}
The first data set (henceforward referred to as set A) consists of nine separate
spectra observed between 7th and 9th of January in 2002. The second set of 6
spectra (B) was obtained on January 13th in 2002. Prior to further data processing
the reduced spectra are reviewed in detail.
\subsubsection{Flux error level}
The given error in flux of all 15 spectra was tested against the zero level
noise in saturated areas.
A broad region of saturated absorption is available near $3913 \AA$ in the
observers frame.
Statistical analysis revealed a variance corresponding to $\sim$120\% of the given
error on average for the 15 spectra. For each of the spectra the calculated
correction factor was applied.
\subsubsection{Relative shifts of the 15 spectra}
\label{preprocessing}
Due to variation in barycentric velocities and grating shifts the individual
spectra are subject to small shifts -- commonly on sub-pixel level -- in
wavelength.
To correct for that effect all spectra were interpolated by a polynomial using
{\it Neville's algorithm} to conserve the local flux (see Fig. \ref{sampled}).
The resulting pixel step on average is $1/20$ of the original data. Each
spectrum was compared to the others. For every data point in a spectrum the
pixel with the closest wavelength was taken from a second spectrum. Their
deviation in flux was divided by the quadratic mean of their given errors in
flux. This procedure was carried out for all pixels inside certain selected
wavelength intervals\footnote{Only certain wavelength ranges are taken into
account to avoid areas heavily influenced by cosmics or areas close to overlapping orders.} resulting in a mean
deviation of two spectra. The second spectrum is then shifted against the first
one in steps of $\sim1.5 $ \ma. The run of the discrepancy of two spectra is
of parabolic nature with a minimum at the relative shift with the best
agreement. Fig. \ref{eta} shows the resulting curve with a parabolic fit. In
this exemplary case the second spectrum shows a shift of 6.2 \ma\ in relation to
the reference spectrum.
The clean parabolic shape verifies the approach. All 15 spectra (A+B) are
shifted to their common mean. The average deviation is 2.3 \ma. Section \ref{influence}
illustrates its influence on the data analysis.
\subsubsection{Selection of lines}
The selection of suitable \hhh features for the final analysis is highly
subjective.
As a matter of course all research groups crosschecked their choice of lines for
unresolved blends or saturation effects. The decision whether a line was
excluded due to continuum contamination or not, however, relied mainly on the
expert knowledge of the researcher and was only partially reconfirmed by the
ascertained uncertainty of the final fitting procedure.
This work puts forward a more generic approach adapted to the fact that we have
two distinct observations of the same object.

\begin{figure}[]
\resizebox{\hsize}{!}{\includegraphics[clip=true]{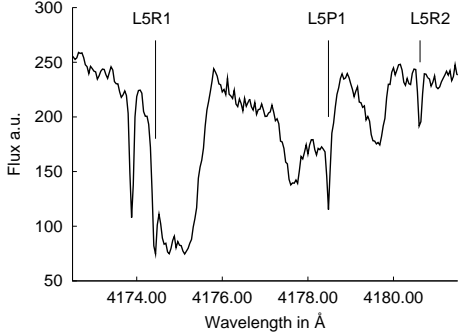}}
\caption{
\footnotesize
Part of the co added spectrum near 4176\AA. The data however, were not co added for the fit.
L5R1 and L5R2 match the 3-$\sigma$ criterion, L5P1 does not.}
\label{3s}
\end{figure}

\begin{figure}[]
\resizebox{\hsize}{!}{\includegraphics[clip=true]{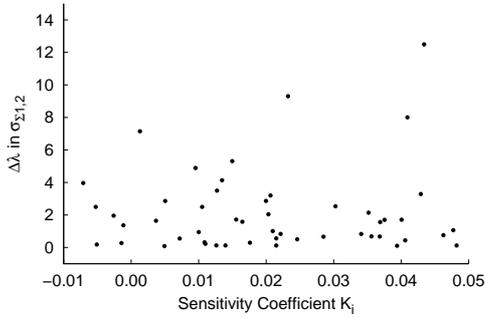}}
\caption{
\footnotesize
Selection of 52 seemingly reasonable lines to be fitted separately for each
dataset
of 9 and 6 exposures, respectively. Their absolute offset
$\Delta\lambda_{\mathrm{effective}}$ to each other is expressed in relation to
their combined error given by the fit (see Eq.\ref{eq1}).
}
\label{compare}
\end{figure}

A selection of 52 lines is fitted separately for each dataset
of 9 (A) and 6 (B) exposures, respectively. In this selection merely blends
readily identifiable or emerging from equivalence width analysis are
excluded.
Each rotational level is fitted with conjoined line parameters except for
redshift naturally. The data are not co added but analyzed simultaneously via
the fitting procedure introduced by \cite{Quast05}.
For each of the 52 lines there are two resulting fitted redshifts or observed
wavelengths, respectively, with their error estimates.
To avoid false confidence, the single lines are not judged by their error
estimate but by their difference in wavelength between the two data sets in relation to the combined 
error estimate. Fig. \ref{compare} shows this dependency.
The absolute offset $\Delta\mathrm{\lambda_{\mathrm{effective}}}$ to each other is expressed in 
relation to their combined error given by the fit:
\begin{equation}
\Delta\lambda_{\sigma_{\Sigma 1,2}} = \frac{\Delta\lambda_{\mathrm{effective}}}{\sqrt{\sigma^2_{\lambda_1}+\sigma_{\lambda_2}^2}}.
\label{eq1}
\end{equation}
Fig. \ref{compare} reveals notable discrepancies between the two datasets, the disagreement is partially
exceeding the 5-$\sigma$ level\footnote{Lines fitted with seemingly high precision and thus a low error
reach higher offsets than lines with a larger estimated error at the same discrepancy in
 $\lambda_{\mathrm{obs}}$. Clearly the lower error estimates merely reflects the statistical
quality of the fit, not the true value of the specific line position.}
Since the fitting routine is known to provide proper error estimates
\citep{Quast05} and \citep{Wendt08}, the dominating source of error in the determination
of line positions is due to systematic errors. This result indicates calibration issues
of some significance at this level of precision. The comparison of two independent observation runs
reveals a source of error that can not be estimated by the statistical quality of the fit alone.
For further analysis only lines that differ by less than 3 $\sigma$ are taken into account.
This criterion is met by 36 lines. Fig. \ref{3s} shows three \hhh features corresponding to the 
transitions L5R1, L5P1, L5R2. All have similar sensitivity towards changes in $\mu$. L5P1 fails the
applied self consistency check between the two data sets and is excluded in the further analysis.

\section{Results}
\begin{figure}[]
\resizebox{\hsize}{!}{\includegraphics[clip=true]{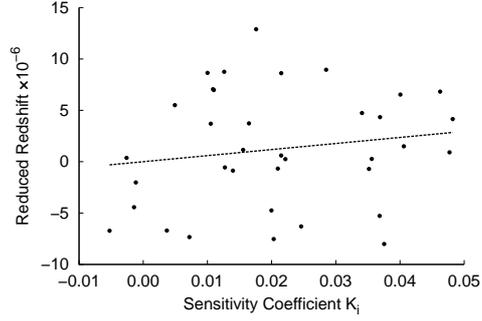}}
\caption{
\footnotesize
The unweighted fit for Q0347 corresponds to $\Delta\mu/\mu = (15 \pm 16) \times
10^{-6}$.
Note, that at such a high scatter $z_{Ki=0}$ differs from $\bar{z}$ by less than
1 $\sigma_\mathrm{z}$.}
\label{result}
\end{figure}

For the final analysis the selected 36 lines are fitted in all 15 shifted, error-scaled 
spectra simultaneously. The result of an unweighted linear fit corresponds to
 $\Delta\mu/\mu = (15 \pm 16) \times 10^{-6}$ over the look-back time of $\sim11.5$ Gyr for 
 $z_{\mathrm{abs}}=3.025$. Fig. \ref{result} shows the resulting plot. Note the comparable large
scatter in determined redshift when combining the two independent observations.
The approach to apply an unweighted fit is a consequence of the unknown nature of the
prominent systematics. Uncertainties in wavelength calibration can not be expressed directly as
an individual error per line.
For this work the calibration errors and the influence of unresolved blends are assumed to be
dominant in comparison to individual fitting uncertainties per feature.
For the following analysis the same error is adopted for each line.
With an uncertainty in redshift of $1\times 10^{-6}$ we obtain:
$\Delta\mu/\mu = (15 \pm 6) \times 10^{-6}$.\\
However the goodness-of-fit is below 1 ppm and is not self consistent.
Judging by that and Fig. \ref{result}, a reasonable error in observed redshift should
at least be in the order of $5\times 10^{-6}$. The weighted fit gives:
$\Delta\mu/\mu = (15 \pm 14) \times 10^{-6}$.\\
\subsection{Uncertainties in the sensitivity coefficients}
At the current level of precision, the influence of uncertainties in the sensitivity 
coefficients $K_i$ is minimal. It will be of increasing importance though when
wavelength calibration can be improved by pedantic demands on future observations and
eventually Laser Frequency Comb calibration will allow for practically arbitrary precision
and uncertainties in the calculations  of sensitivities will play a role.
Commonly the weighted fits neglects the error in $K_i$.
The $\chi^2$ merit function for the case of a straight-line fit with errors in both coordinates
is given by:
\begin{equation}
\chi^2(a,b) = \sum^{N-1}_{i=0}\frac{(y_i-a-bx_i)^2}{\sigma^2_{yi}+b^2\sigma^2_{xi}}
\end{equation}
and can be solved numerically with valid approximations \citep{Lybanon84}.
At the current level even an error in $K_i$ of about 10\% has merely impact on the error estimate
in the order of $10^{-6}$.\\
Alternatively the uncertainties in $K_i$ can be translated into an uncertainty in redshift via
the previously fitted slope. The results of this ansatz are similar to the fit with errors in both
coordinates and in general this is simpler to implement.\\
Another possibility is to apply a gaussian error to each sensitivity coefficient and redo the
normal fit a few dozen times with alternating variations in $K_i$.
Again, the influence on the error-estimate is in the order of 1 ppm.\\
The different approaches to the fit allow to estimate its overall robustness as well.
\subsection{Individual line pairs}
\begin{figure}[]
\resizebox{\hsize}{!}{\includegraphics[clip=true]{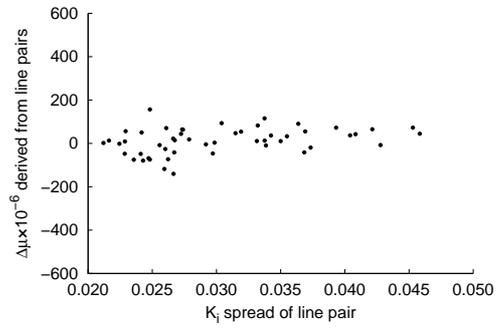}}
\caption{
\footnotesize
$\Delta\mu/\mu$ derived from individual line pairs (52) which are separated by less
than 110\AA\  and show a difference in sensitivity of more then $0.02$.\newline
 The median corresponds to $\Delta\mu/\mu = 13\times10^{-6}$.
}
\label{dmu}
\end{figure}

 $\Delta\mu/\mu$ can also be obtained by using merely two lines that show different sensitivity towards changes
in the proton-to-electron mass ratio. Another criterion is their separation in the wavelength frame to
avoid pairs of lines from different ends of the spectrum. Several tests showed that a separation of
$\Delta\lambda \leq 110 \AA$ and a range of sensitivity coefficients $K_1-K_2 \geq 0.02$ produces stable
results that do not change any further with more stringent criteria. Note, that pairs spread across
two neighboring orders ($\sim50 \AA$) show no striking deviations.
Fig.\ref{dmu} shows the different values for $\Delta\mu/\mu$ derived from 52 line pairs that match the
aforementioned criteria. Note, that a single observed line contributes to multiple pairs. The gain in
statistical significance by this sorting is limited as pointed out by \cite{Molaro08}.
The large scatter favors the median as result which produces $\Delta\mu/\mu = 13 \times 10^{-6}$.
The scatter is then related to uncertainties in the wavelength determination which is mostly due to calibration
errors. The standard error is $8 \times 10^{-6}$.
\subsection{Influence of the preprocessing} \label{influence}
\begin{figure}[]
\resizebox{\hsize}{!}{\includegraphics[clip=true]{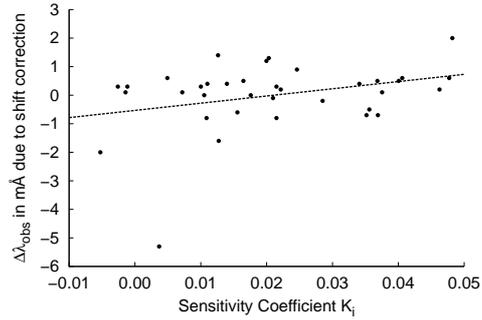}}
\caption{
\footnotesize
Variation in fitted positions for all lines with and without initial
correction for shifts in between the 15 spectra. The slope of the fit
is dominated by three lines. 
}
\label{shift_influence}
\end{figure}

Section \ref{preprocessing} describes the initial shift to a common mean of all 15 spectra.
The complete analysis was redone with error-scaled but unshifted spectra and the ascertained line positions
of both runs compared. Fig. \ref{shift_influence} shows the difference for each \hhh line in m\AA\ over
the corresponding sensitivity coefficients $K_i$. The plotted line is a straight fit. Clearly the slope
is dominated by three individual lines whose fitted centroids shifted up to $5.5$ m\AA\ due to the preprocessing.
These three lines in particular produce a trend towards variation in $\mu$ when grating shifts and other effects
are not taken into account. This single-sided trend probably occurred by mere chance but at such low statistics
it influences the final result.

\begin{acknowledgements}
We are thankful for the support from the Collaborative Research Centre 676 and for
helpful discussions on this topic with S.A. Levshakov, P. Petitjean and M.G. Kozlov.
\end{acknowledgements}

\bibliographystyle{aa}

\end{document}